\documentclass[aps, prl, superscriptaddress, nofootinbib, 
               amsmath, amssymb, twocolumn,
               preprintnumbers, showpacs,
               raggedbottom,
               floatfix,
               ]{revtex4-1}
\usepackage{here}
\usepackage[utf8]{inputenc}
\usepackage[arXiv]{my_paper}                     
\usepackage{my_acronyms}                         

\newcommand{\h}{l_{\text{coh}}}
\let\Im\relax
\DeclareMathOperator{\Im}{Im}

\begin{document}

\title{Quantized Superfluid Vortex Rings in the Unitary Fermi Gas}

\def\email#1{}
\author{Aurel Bulgac}
\email{bulgac@uw.edu}
\affiliation{Department of Physics, University of Washington, Seattle,
  Washington 98195--1560, USA}


\author{Michael McNeil Forbes}
\email{mforbes@alum.mit.edu}
\affiliation{Institute for Nuclear Theory, University of Washington,
  Seattle, Washington 98195--1550, USA}
\affiliation{Department of Physics, University of Washington, Seattle,
  Washington 98195--1560, USA}
\affiliation{Department of Physics \& Astronomy, Washington State University,
  Pullman, Washington 99164--2814, USA}

\author{Michelle M. Kelley}
\email{kelley19@illinois.edu}
\affiliation{Department of Physics, University of Illinois at Urbana/Champaign,
  Il 61801-3080} 

\author{Kenneth J. Roche}
\email{k8r@u.washington.edu}
\affiliation{Pacific Northwest National Laboratory,
  Richland, Washington 99352, USA}
\affiliation{Department of Physics, University of Washington, Seattle,
  Washington 98195--1560, USA}

\author{Gabriel Wlaz\l{}owski}
\email{gabrielw@if.pw.edu.pl}
\affiliation{Faculty of Physics, Warsaw University of Technology,
  Ulica Koszykowa 75, 00-662 Warsaw, Poland}
\affiliation{Department of Physics, University of Washington, Seattle,
  Washington 98195--1560, USA}

\date{\today}

\begin{abstract}\noindent
  In a recent article, Yefsah \textit{et al.} [Nature
  \textbf{499}, 426 (2013)] report the observation of an unusual excitation in
  an elongated harmonically trapped unitary Fermi gas. After phase imprinting a
  domain wall, they observe oscillations almost an order of magnitude slower
  than predicted by any theory of domain walls which they interpret as a ``heavy
  soliton'' of inertial mass some 200 times larger than the free fermion mass or
  50 times larger than expected for a domain wall.  We present compelling
  evidence that this ``soliton'' is instead a quantized vortex ring by showing
  that the main aspects of the experiment can be naturally explained within the
  framework of time-dependent superfluid \glspl{DFT}.
\end{abstract}
\preprint{INT-PUB-13-021}
\preprint{NT@UW-13-21}
\pacs{
67.85.Lm,  
67.85.De,  
03.75.Ss,  
03.75.Kk,  
67.85.-d,  
05.30.Fk,  
}
\maketitle
\glsresetall
\glsunset{GPU}
\glsunset{MIT}
\setlength\marginparwidth{40pt}

\lettrine{C}{ollective modes} in the form of topological and dynamical defects
-- solitons, vortices, vortex rings, etc.\@ -- embody the emergence of
non-trivial collective dynamics from microscopic degrees of freedom, and provide
a challenge for many-body theories from cold atoms through electronic
superconductors to nuclei and neutron stars.  The \gls{UFG} provides an ideal
strongly interacting system for measuring and testing collective modes where
controlled experiments and theoretical techniques are starting to
converge~\cite{Zwerger:2011}. A handful of predicted collective modes have been
directly observed, including collective oscillations of harmonically trapped
gases~\cite{Bartenstein;Altmeyer;Riedl;Jochim;Chin...:2004-05, *Altmeyer:2007a,
  *Wright:2007, *Riedl:2008,
  Kinast;Hemmer;Gehm;Turlapov;Thomas:2004-04,*Kinast:2004}, higher-nodal
collective modes~\cite{Guajardo:2013}, scissor modes~\cite{Wright:2007},
quantized vortices and vortex lattices~\cite{ZA-SSSK:2005lr}, shock
waves~\cite{Joseph:2011}, and phonons (speed of sound~\cite{Joseph:2007},
critical velocity~\cite{Miller:2007}, and first and second sound~\cite{Tey:2013,
  *Sidorenkov:2013, Guajardo:2013}). Other modes, such as the Higgs
mode~\cite{Bulgac:2009b, Scott:2012}, vortex rings~\cite{Bulgac:2011b}, and
domain walls~\cite{Scott:2011,Antezza:2007, Spuntarelli:2011, Liao:2011,
  Bulgac:2011c}, have been demonstrated in simulations, but await direct
observation.  In this paper, we discuss the objects observed
in~\cite{Yefsah:2013}: they interpret these as ``heavy solitons''; we show them
to be vortex rings.

\paragraph{Experimental Puzzle: Slowly moving ``solitons''}
The recent \gls{MIT} experiment~\cite{Yefsah:2013} measures a slowly
moving ``soliton'' produced by a sharp spatially delineated phase imprint on an
ultracold cloud of some $10^5$ \ce{^{6}Li} atoms in an elongated harmonic trap.
These ``solitons'' cannot be resolved \textit{in situ}, but appear after a
specific time-of-flight expansion procedure of that includes a rapid ramp of the
interaction strength which is controlled through a Feshbach resonance by an
external magnetic field.  In particular they note that a certain minimum field
$B_\text{min}<700$~G is required to resolve the ``solitons'' (discussed in their
supplementary material). From the images, they extract the period of
oscillation, and find that it increases as  the
inverse trap aspect ratio $1/\lambda$ and the magnetic field $B$ are 
increased.  Increasing the
temperature, they observe ``anti-dampening'' whereby the amplitude of the
oscillation increases with time. The authors interpret these results as the
observation of a ``heavy soliton'' with a mass ``more than 50 times larger than
the theoretically predicted value'' and ``200 times their bare
mass.\kern-2pt''

\paragraph{Topological objects in the BEC-BCS crossover}
Superfluids are characterized by a complex-valued order parameter $\Psi$ that
describes the condensate wavefunction in \glspl{BEC} and the Cooper pair
condensate in fermionic \gls{BCS} superfluids.  The superfluid ground state
picks a coherence overall phase of the complex order parameter, spontaneously
breaking the original $U(1)$ phase symmetry of the theory. Sound waves manifest
as fluctuations in this coherent phase (phonons or Nambu-Goldstone modes).
Landau's original argument for $^4$He superfluidity posits a kinematical
critical flow velocity $v_c$ below which neither pair-breaking nor sound
excitations can be generated.  This argument is spoiled by the generation of
topologically stable excitations that can nucleate at the edge of the fluid,
lowering the $v_c$.  The dynamics of these topological excitations and their
interactions are at the heart of quantum turbulence studies~\cite{Vinen:2002,
  *Vinen:2006, *Vinen:2010, *Skrbek:2011, *Tsubota:2008, *Tsubota:2013,
  *Paoletti:2011}.

The single-valued order parameter admits several topologically stable objects in
three dimensions. Domain walls separate regions of different phases while
vortices correspond to the phase winding around a line along which the order
parameter vanishes.  In bosonic theories (\gls{BEC} limit), the number density
$n \propto \abs{\Psi}^2$ vanishes in the core of vortices and in stationary
domain walls, giving these objects a ``negative mass.\kern-2pt'' For fermions,
while the complex order parameter has a similar behavior, the relationship $n
\propto \abs{\Psi}^2$ breaks down, with the interpretation that the core of the
topological defects are filled with ``normal'' fluid, but at unitarity the
number density depletion is still substantial~\cite{BY:2003, Yu:2003a,
  Bulgac:2011b}.  A manifestation of this negative mass is that the amplitude of
oscillation in a trap will increase as energy is lost.  This ``anti-damping'' is
seen in the experiment~\cite{Yefsah:2013}.

Domain walls (often referred to as solitons) are topologically stable in one
dimension.  Their thickness is set by the coherence length $\h$ and thus have a
negative effective mass ($-M_{\textit{DW}}$) due to the density depletion
$M_{\textit{DW}} = mN_{\textit{DW}}$ where $N_{\textit{DW}} \sim n \pi R^2 \h$ is the
depletion for a gas cloud of number density $n$ in a trap of radius $R$.  In the
unitary limit, all scales are set by the Fermi wavevector $k_F$ with $n =
k_F^3/3\pi^2$ and $\h \sim k_F^{-1}$ and thus, $M_{\textit{DW}} \sim k_F^2 R^2 m$
is much larger than the mass $m$ of a single fermion.  In quantum mechanics, the
dynamics of heavy objects is generally well approximated by classical equations
of motion.  For domain walls, both kinetic and potential energies are localized
on the wall, thus the same mass $M_{\textit{DW}}$ enters both the kinetic and
potential terms and one expects the oscillation period $T$ to be comparable to
the natural axial period $T_z$ of the trapping potential. This is confirmed in
\gls{BEC} experiments~\cite{Becker:2008, *Weller:2008} where $T \approx
\sqrt{2}T_z$ and by fermionic simulations~\cite{Scott:2011} where $T \approx
\sqrt{3}T_z$.

In contrast, vortex rings~\cite{Rayfield:1964}, which also occur in
classical fluids~\cite{Lamb:1945, *Saffman:1992}, have very different dynamics.
In infinite media, for example, with logarithmic accuracy, large rings ($R \gg
\h \sim k_F^{-1}$) have linear momentum $p \sim m n \kappa \pi R^2$, dispersion
$\varepsilon(p)$, and speed $v = \d{\varepsilon(p)}/\d{p}$~\cite{Roberts:1970,
  *Barenghi:2009}:
\begin{subequations}
  \label{eq:vortex_ring}
  \begin{equation}
    \varepsilon \sim \frac{m n  \kappa^2 R}{2} \ln\frac{R}{\h}, \quad
    v \sim \frac{\kappa}{4\pi R}\ln\frac{R}{\h}
    \label{eq:1}
  \end{equation}
  where $\kappa$ is the circulation. Their speed $v\propto \ln p/\sqrt{p}$ thus
  \emph{decreases} as the momentum, kinetic energy, and radius increase.  Unlike
  for domain walls, their inertial mass $M_{I} = F/\dot{v} \sim m n \kappa
  8\pi^2 R^3/\ln(R/\h)$, (where $F=\dot{p}$ is the force), differs from the
  effective mass due to the density depletion $M_{\textit{VR}} = m
  N_{\textit{VR}} \sim m n 2\pi^2 R\h^2$ and the period of oscillation can
  receive a significant enhancement
  \begin{gather}
    \label{eq:ring_period}
    \frac{T}{T_z} \sim \sqrt{\frac{M_{I}}{M_{\textit{VR}}}} \sim
    \frac{2R/\h}{\sqrt{\ln(R/\h)}}.
  \end{gather}
\end{subequations}
This estimate~\eqref{eq:ring_period} gives only an order of magnitude estimate:
the dynamics of a vortex ring in a finite trap is somewhat more complicated but
can be qualitatively understood. Each element of the ring will experience an
outward buoyant force $\vect{F}_{B}\approx
N_{\textit{VR}}\vect{\nabla}V_{\text{trap}}$ where $V_{\text{trap}} = m
\omega_{\perp}^2(x^2+y^2+z^2/\lambda^2)/2$ (with $\lambda > 1$).  The Magnus
relationship $\vect{F}_{B} = mn(\vect{v} - \vect{v}_s)\times \vect{\kappa}$ will
thus adjust the velocity $\vect{v}$ with two components: one counter to
$\vect{v}_s$ and another that causes the ring to expand and contract near the
ends of the trap.  The velocity $\vect{v}_s$ is the superflow induced by the
phase winding of the rest of the vortex ring on the element, and is parallel to
the $z$-axis of the trap.  In the middle of the trap $z \approx 0$, small rings
($R$ much less than the trap waist $R_\perp$) will experience little buoyant
force and the motion will be dominated by $\vect{v} \approx \vect{v}_s$. Larger
rings, however, will have a smaller $\vect{v}_s$ and larger $\vect{F}_{B}$: at a
critical radius $R_c$, the Magnus effect will cancel $\vect{v}_s$ and the ring
will remain stationary.  Larger rings $R > R_c$ will crawl backwards along the
trap.  Near the ends of the trap, the buoyant force will also cause the rings to
expand at one end and contract at the other.  Thus a vortex ring may oscillate
along the trap as observed in bosons~\cite{Shomroni:2009}.

\paragraph{``Heavy solitons'' are indeed vortex rings}
While a quantitative discussion requires a more complete analysis along the
lines of~\cite{Komineas:2007} or direct simulation as we shall present in a
moment, the order of magnitude of the effect can be estimated from
Eq.~\eqref{eq:ring_period} which is approximately valid for small vortex rings
near the middle of the trap $z \approx 0$. For the experimental parameters,
small rings $R \approx 0.2R_\perp$ (rings with this radius have roughly the same
amplitude as the oscillations seen in the experiment) exhibit periods an order
of magnitude larger than $T_z$, naturally explaining the observations.
Furthermore, as the system is brought into the \gls{BEC} regime, the coherence
length $\h$ grows significantly relative to the fixed ring size $R$, so $T$
naturally becomes smaller, approaching $T_z$.  Finally, in the extreme
\gls{BEC} limit, $\h$ approaches the width of the trap, arresting the snake
instability, and reproducing the theoretical prediction $T \approx \sqrt{2}T_z$
for a domain wall.

\paragraph{Method}
To explain more subtle features of the experiment, like the observed dependence
on aspect ratio, we perform dynamical simulations of trapped unitary fermions
using two formulations of \gls{DFT}.  The first, an \gls{ETF}
model~\cite{Kim:2004a, *Kim:2004, *Salasnich:2008b, *Salasnich:2008,
  *Salasnich:2008E}, is essentially a bosonic theory for the dimer/Cooper-pair
wavefunction $\Psi$.  The dynamics are described by a \gls{NLSEQ} similar to the
\gls{GPE} for bosons
\begin{gather}
  \label{eq:ETF}
  \I \hbar \pdiff{\Psi}{t} = -\frac{\hbar^2}{4m}\vect{\nabla}^2\Psi+
  2\frac{\partial \mathcal{E}_{h}(n,a)}{\partial n}\Psi + 2V_\text{ext}\Psi
\end{gather}
where arguments $\vect{x}$ and $t$ have been suppressed, $n = 2\abs{\Psi}^2$ is
the fermion number density, and $\mathcal{E}_{h}(n,a)$ is the energy-density of
the homogeneous gas with density $n$ and (adjustable) scattering length $a$ fit
to the equation-of-state in the \gls{BEC}-\gls{BCS} crossover.  This simplified
\gls{DFT} is equivalent to zero-temperature quantum hydrodynamics (including the
so called quantum pressure term), and we shall use this to model the
time-of-flight expansion/imaging procedure of the experiment.  While
computationally attractive, this formulation has some physical drawbacks. In
particular, it models only the superfluid portion of the cloud: physics
associated with the normal state is missing.  As a result, a vanishing order
parameter $\Psi = 0$ implies a vanishing density $n = 0$. This tends to
overestimate the density contrast in the core of defects and leads to the same
domain wall motion $T\approx\sqrt{2}T_z$ as the harmonically trapped \gls{GPE}.
There is also no mechanism for the superfluid to transfer energy to the normal
component, which inhibits the relaxation of rotating systems into a regular
vortex lattice, and prevents Eq.~\eqref{eq:ETF} from being used to simulate the
preparation of the experiment as the initial sound waves generated by the phase
imprint never dampen, and the generated vortex rings rapidly decay.

To address these issues, we also simulate a time-dependent extension of
\gls{DFT} to superfluid systems -- the \gls{TDSLDA} -- where the dynamical
evolution is described by equations for the quasiparticle wavefunctions
$(u_k,v_k)$
\begin{subequations}\label{eq:SLDA}
  \begin{equation}
    \I \hbar \pdiff{}{t}
    \begin{pmatrix}
      u_k \\ 
      v_k
    \end{pmatrix}
    =
    \begin{pmatrix}
      h  & \Delta \\
      \Delta^* & -h
    \end{pmatrix}
    \begin{pmatrix}
      u_k \\
      v_k
    \end{pmatrix},
    \label{eq:4}
  \end{equation}
  where $h=\delta \mathcal{E}/\delta n$ and $\Delta=\delta \mathcal{E}/\delta
  \nu^*$ ($\nu$ is the anomalous density)~\cite{Bulgac:2007a, *Bulgac:2011,
    *Bulgac:2013b}.  This is similar in form to the \gls{BdG} mean-field
  theory~\cite{Antezza:2007,Scott:2011, Spuntarelli:2011, Liao:2011}, but
  includes a self-energy contribution $\beta$ and effective mass parameter
  $\alpha$ neglected in the \gls{BdG}:
  \begin{equation}
    h = \frac{ \delta \mathcal{E} }{ \delta n } 
    = \alpha \frac{-\hbar^2\vect{\nabla}^2}{2m} 
    +  \beta \frac{(3\pi^2n)^{2/3}}{2}  
    - \frac{\abs{\Delta}^2}{3\gamma n^{2/3}}.
  \end{equation}
\end{subequations}
These additional terms allows the \gls{TDSLDA} to quantitatively match all
experimentally measured and numerically calculated properties of homogeneous
systems in finite and infinite boxes~\cite{Forbes:2012}: adjusting $\alpha$,
$\beta$, and $\gamma$ allows one to consistently characterize the energy per
particle, pairing gap, and quasiparticle spectrum obtained from \ac{QMC}
calculations of the homogeneous infinite system. (Note: If $\alpha \neq 1$, one
must include additional terms to restore Galilean covariance as discussed
in~\cite{Bulgac:2007a, *Bulgac:2011, *Bulgac:2013b, EPAPS:aps}: we avoid this
complication by setting $\alpha=1$ instead of $\alpha \approx 1.1$ while
adjusting $\beta$ and $\gamma$ to reproduce the energy per particle and pairing
gap.)  Simulating Eqs.~\eqref{eq:SLDA} for three-dimensional systems represents
a serious computational challenge that effectively utilizes the largest
supercomputers available, so we use this only to verify that stable vortex rings
are generically produced from the phase-imprint procedure, and use the
\gls{ETF}~\eqref{eq:ETF} to model the experimental systems.

\begin{figure}
  \includegraphics[width=\columnwidth]{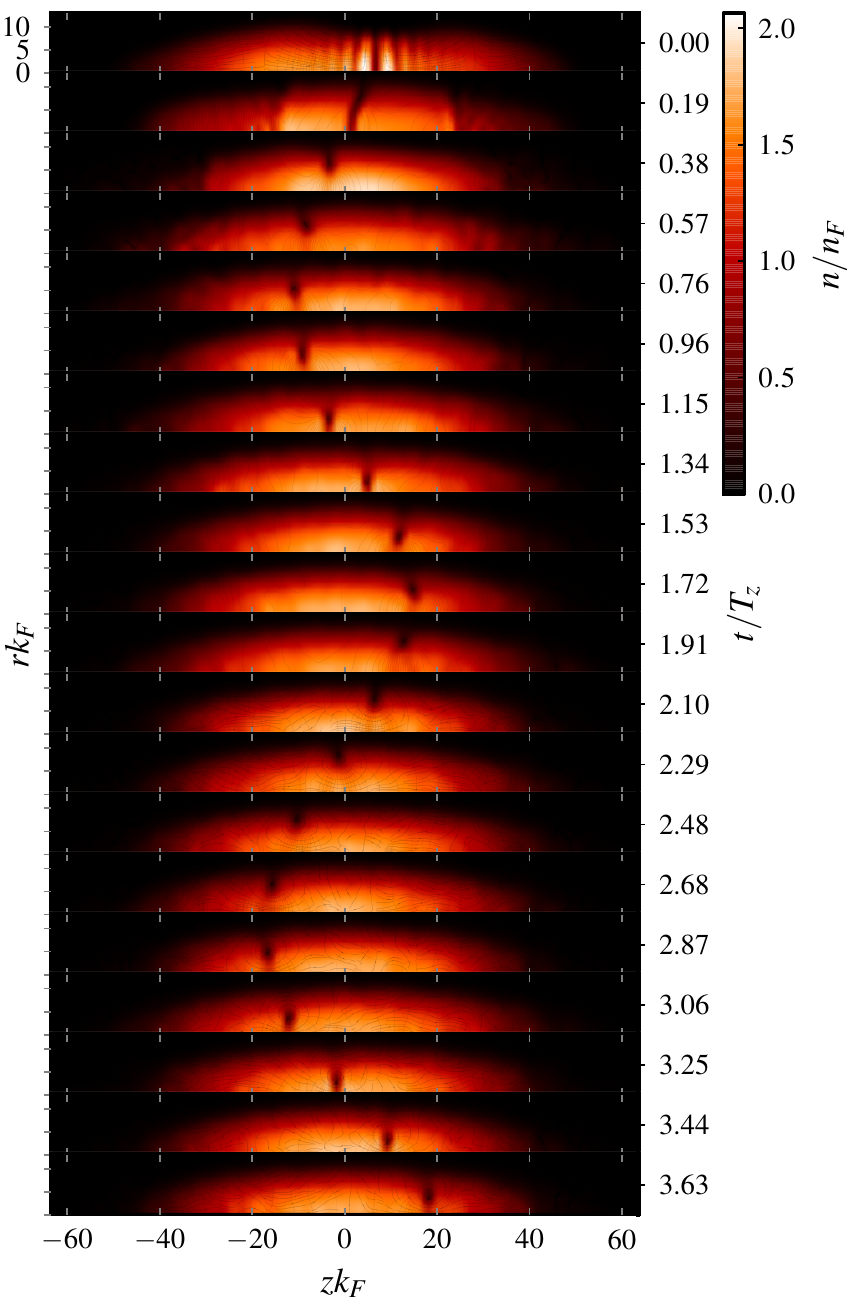}  
  \caption{(color online) \label{fig:vortexring_2}%
    Oscillations of a vortex ring in an elongated harmonic trap.  Simulated with
    the \gls{TDSLDA} on a $32\times32\times128$ lattice for a cloud with 560
    particles. We evolve about $10^5$ wavefunctions in real time using a
    symplectic split-operator integrator that respects time-reversal invariance
    using hundreds of \acp{GPU} on the Titan supercomputer~\cite{titan}.  More
    details and several movies may be found in~\cite{EPAPS:aps}}
\end{figure}

\begin{figure}
  \includegraphics[width=\columnwidth]{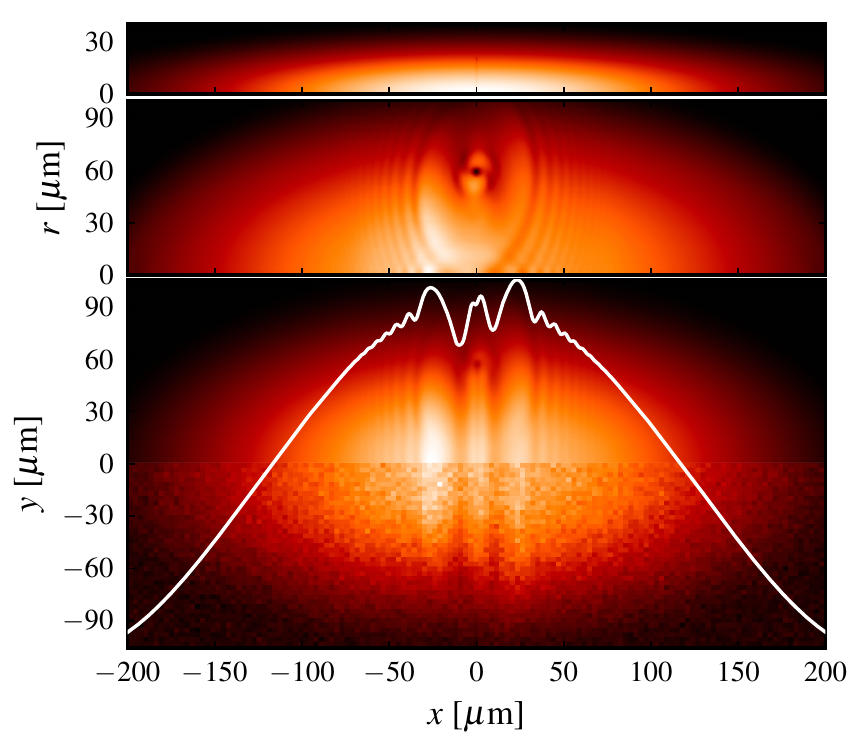}
  \caption{(color online) \label{fig:imag}%
    Demonstration of the imaging procedure. The top plot shows a slice of the
    density through the upper-half core of the trap before expansion: the vortex
    ring is barely visible at $z=0$.  Below is a slice through the upper-half
    core after ramping to $B_{\text{min}}=580$~G and letting the cloud expand as
    discussed~\cite{Yefsah:2013}.  The lower plot shows the integrated \mysc{2D}
    density $\int \!\d{x}\, n(x,y,z)$ and the integrated \mysc{1D} density $\int
    \!\d{x}\d{y} \,n(x,y,z)$ (white curve).  The lower half of the image has
    added Gaussian noise with a 3\% density variation and is coarse-grained on a
    3~$\mu$m scale to simulate the experimental imaging procedure, clearly
    demonstrating that vortex rings appear as solitons. (Densities are scaled by
    maximum value for better contrast.) For $B_{\text{min}}>700$~G, the density
    contrast is reduced below the experimental signal-to-noise ratio.  See the
    supplementary information~\cite{EPAPS:aps} for details and for movies.}
\end{figure}

\paragraph{Results}
Following the preparation procedure outlined in~\cite{Yefsah:2013}, we phase
imprint a domain wall on harmonically trapped clouds and follow the evolution
using the~\gls{TDSLDA}. (Details are presented in~\cite{EPAPS:aps}.)  For
sufficiently large clouds, the domain wall quickly decays into an oscillating
vortex ring. Fig.~\ref{fig:vortexring_2} shows the motion as the ring initially
crawls along the outside of the trap and a smaller ring bounces
back. Computational limitations restrict us to relatively small systems and
these simulations are quite close to the onset of the snake instability.
Nevertheless, the period seen in Fig.~\ref{fig:vortexring_2} is comparable to
our estimate Eq.~\eqref{eq:ring_period}. Finally, we note an anti-damping
similar to that seen at higher temperatures in~\cite{Yefsah:2013}. This is
explained by the small heat-capacity of our simulated system: the residual sound
waves induced by the phase correspond roughly to a finite temperature.

For larger clouds we use the \gls{ETF}~\eqref{eq:ETF}.  As expected, the initial
preparation phase cannot be reliably reproduced: the sound waves generated by
the imprint do not dissipate, and the resulting vortex ring decays within a few
oscillations.  Stable vortex rings can be produced, however, by ``cooling'' an
imprinted phase pattern with imaginary time evolution.  As shown in the
supplementary information~\cite{EPAPS:aps}, these vortex rings reproduce the
qualitative behaviors observed in the \gls{MIT} experiment~\cite{Yefsah:2013}.
In particular, the period is an order of magnitude larger than expected for
domain walls and increases by similar amounts as the aspect ratio is reduced as
shown in Table \ref{tab:ETF_aspect}.  The period also scales toward the domain
wall results $\sqrt{2}T_z$ toward the \gls{BEC} limit and exhibits anti-damping
decays in the presence of phonon excitations. (These phonons mock up
fluctuations, but do not faithfully simulate a thermal ensemble.)  A
quantitative comparison is marred by the lack of a normal component occupying
the core of the vortex. However, when comparing the \gls{ETF} with the
\gls{TDSLDA} simulations, we find that this is fairly consistently characterized
by an overall increase in periods by a factor of about $1.8$ -- somewhat larger
but similar to the factor of $\approx \sqrt{3/2}$ seen when comparing the period
of fermionic to bosonic domain walls in quasi-\mysc{1D} environments.  We are
confident that a realistic \gls{TDSLDA} simulation would closely mimic the
experiment, and enforcing quantitative agreement would help further constrain
the \gls{TDSLDA} functional.

\begin{table}[h]
  \caption{\label{tab:ETF_aspect}%
    Dependence of the oscillation period on aspect ratio for a vortex ring
    imprinted with $R_0=0.30~ R_\perp$ at resonance.  Note that the \gls*{ETF}
    consistently underestimates the period by about a factor of $0.56$.} 
  \begin{ruledtabular}
    \begin{tabular}{lll}
      Aspect Ratio & \gls*{ETF} Period & Observed Period~\cite{Yefsah:2013}\\
      \hline         
      $\lambda=\hphantom{1}3.3$ & $T=9.9~T_z$ & $T = 18(2)T_z$\\
      $\lambda=\hphantom{1}6.2$ & $T=8.4~T_z$ & $T = 14(2)T_z$\\
      $\lambda=15$ &$T=6.7~T_z$ & $T = 12(2)T_z$\\
    \end{tabular}
  \end{ruledtabular}
\end{table}

The puzzle provided by the imaging procedure remains: can a vortex ring look
like a planar soliton after imaging?  The answer, yes, is demonstrated in
Fig.~\ref{fig:imag} and in~\cite{EPAPS:aps}.  The imaging procedure includes a
rapid ramp of the magnetic field to the \gls{BEC} side of the crossover where
the coherence length becomes much larger, but the equation of state becomes
softer.  This rapid-ramp procedure followed by expansion produces something akin
to a shock wave~\cite{Joseph:2011,Bulgac:2011c} that manifests itself as a
planar soliton upon imaging. Our simulations confirm the somewhat subtle
experimental observation that sufficient ramping below $B_{\text{min}} < 700$~G
is required to observe a signal, and explains both the thickness of the
``soliton'' and the amplitude of the integrated density fluctuations observed in
the experiment~\cite{Yefsah:2013}. A slight difference remains between This
deficiency of the \gls{ETF} can also explain a quantitative difference between
the density fringe pattern seen in the integrated \mysc{1D} density
Fig.~\ref{fig:imag} compared with those seen in experiment~\cite{Yefsah:2013},
the latter having a minimum in the center where the \gls{ETF} has a peak. As
shown in the movies of the expansion~\cite{EPAPS:aps}, this feature results from
the motion of shock-waves formed during the expansion, the speed of which is
incorrectly predicted by the \gls{ETF}.

We have shown that the puzzling report of ``heavy solitons'' in fermionic
superfluids~\cite{Yefsah:2013}, which appear to exhibit an effective mass some
fifty times larger than predicted by theory of dynamics of a domain wall, can be
naturally explained in terms of vortex rings.  Using a \mysc{3D} simulation of
the \gls{TDSLDA}, we validate the picture that, in large enough traps, imprinted
domain walls generically evolve into vortex rings through an axially-symmetric
``snake instability.\kern-2pt'' The estimate Eq.~\eqref{eq:ring_period} shows
that these rings can have large periods at unitarity, that decreases toward the
\gls{BEC} regime, and explicit simulations using the \gls{ETF} verify the
dependence of the period on the aspect ratio.  Finally, the \gls{ETF}
demonstrates that, through the expansion/imaging process employed to resolve the
objects, vortex rings manifest as large planar objects with an observable
density contrast only if the magnetic field is ramped to $B_{\text{min}} <
700$~G, in quantitative agreement with the observations. We have thus verified
virtually all aspects of the experiment~\cite{Yefsah:2013}, including the
elaborate imaging protocol, thereby validating the use of the \gls{TDSLDA} and
\gls{ETF} theories for dynamical simulations including topological defects, and
resolving the mystery of ``heavy solitons'' as vortex rings.

\acknowledgments
\providecommand{\MMFGRANT}{\MakeUppercase{de-fg02-00er41132}}

\liningnums{\small
We acknowledge support under U.S. Department of Energy (DoE) Grant
Nos. DE-FG02-97ER41014 and \MMFGRANT.  M.M.K. acknowledges the support provided
by an REU NSF fellowship. G.W.\@ acknowledges the Polish Ministry of Science for
the support under Contract No.\@ N\,N202\,128439, within the program ``Mobility
Plus \!-\! I edition'' under Contract No.\@ 628/MOB/2011/0, and the Polish
National Science Center (NCN) decision No.\@ DEC-2013/08/A/ST3/00708.  Some of
the calculations reported here have been performed at the University of
Washington Hyak cluster funded by the NSF MRI Grant No.\@ PHY-0922770. This
research also used resources of the National Center for Computational Sciences
at Oak Ridge National Laboratory, which is supported by the Office of Science of
the DoE under Contract DE-AC05-00OR22725. We thank R.\@ Sharma and M.\@
Zwierlein for discussions.}


%
%
%
%
%
\clearpage
\newpage
\setcounter{page}{1}


\section{Supplementary Information}
\begin{figure}[tp]
  \includegraphics[width=\columnwidth]{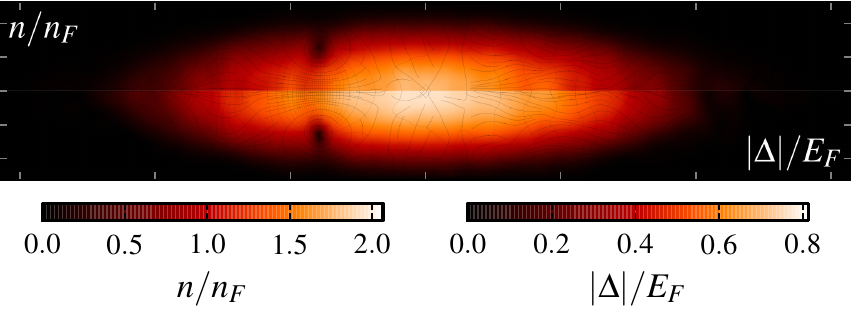}  
  \caption{(color online) \label{fig:single_frame}%
    The shading (color) contours show slice through the core $x=0$ the density
    (top) and magnitude of the order parameter (bottom).  Contours of constant
    phase $\phi = \arg \Delta$ are shown as are streamlines and arrows parallel
    to the superfluid velocity $\vect{v} = \vect{\nabla}\phi$.  A higher
    resolution figure may be found in~\cite{EPAPS:aps}}
\end{figure}

\noindent
The essential feature of the domain walls imprinted in the
experiment~\cite{Yefsah:2013} is that they can decay via a ``snake'' instability
into vortex rings when the radial extent of the trap becomes larger to the
coherence length $R \gg \h$~\cite{Anderson:2001, Brand:2002, Berloff:2002,
  Komineas:2007}.  The formation of an axially aligned vortex rings from a
trapped domain wall is thus almost inevitable: The center of the wall moves
faster than the edges so that the wall bows out along the axis of the trap.  If
the trap is narrow, then the wall maintains integrity (see e.g.\@
Ref.~\cite{Bulgac:2011c}) and one will indeed observe an oscillating domain
wall, but as the trap becomes wider, the bowing out will eventually overwhelm
the domain wall, establish a circulation, and form a vortex ring.  Pinsker
\textit{et al.}~\cite{Pinsker:2013} used this idea to suggest a ``piston
mechanism'' for generating vortex rings, and the \gls{MIT}
experiment~\cite{Yefsah:2013} essentially reproduces this setup. One thus
generically expects a phase imprint to generate vortex rings once the width of
the trap exceeds some critical value.  The detailed structure of one such a ring
from our simulations is down in Fig.~\ref{fig:single_frame} shows a
cross-section of the cloud.

\begin{figure*}%
  \includegraphics[width=\textwidth]{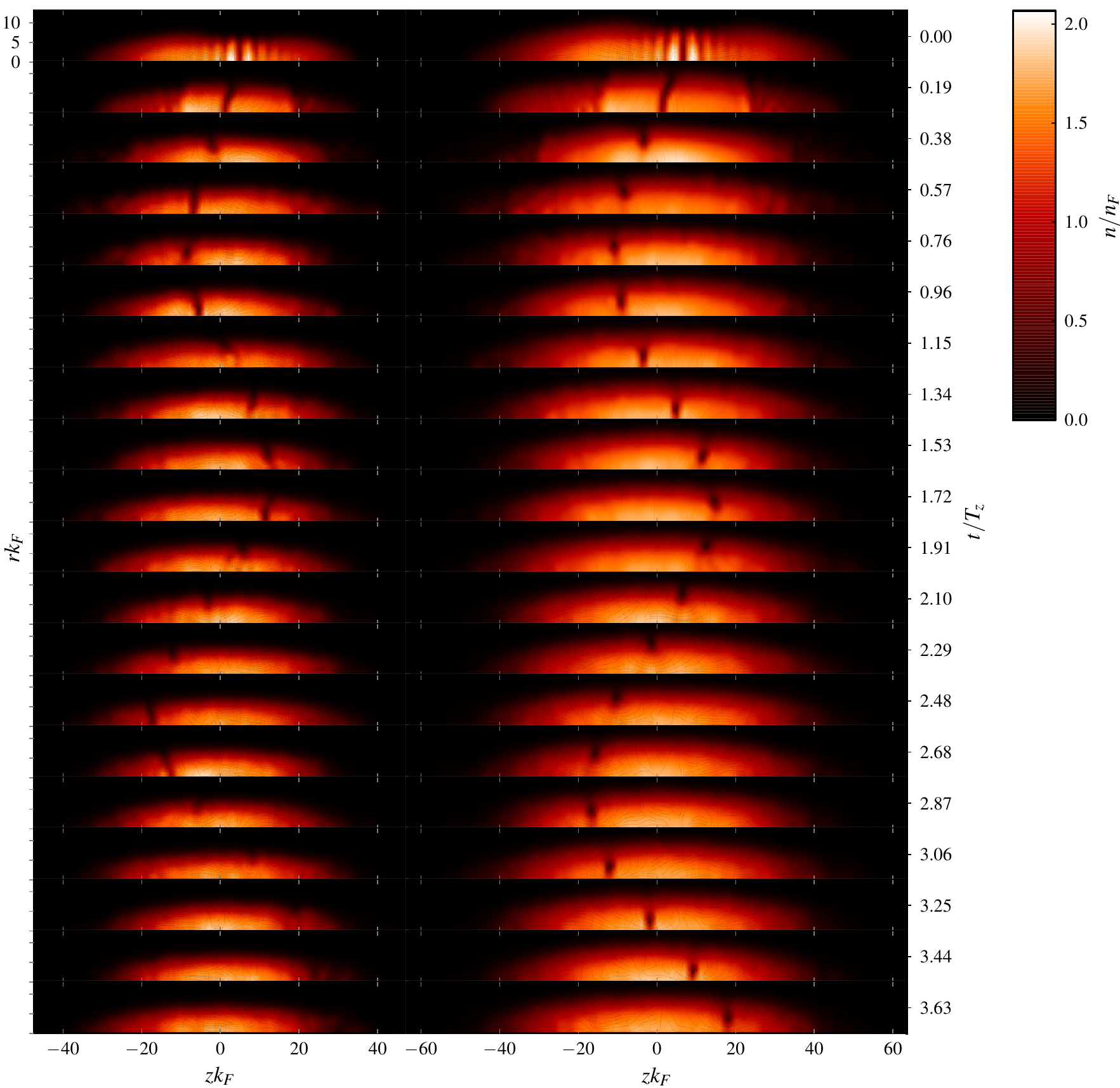}
  \caption{(color online) \label{fig:vortexring}%
    Oscillations of a vortex ring in a harmonic trap on a $24\times 24\times 96$
    lattice (left) and a $32\times32\times128$ lattice (right). We start with a
    cylindrical cloud (not shown, see Ref.~\cite{Bulgac:2013c,EPAPS:aps}) with
    central density $n_F = k_F^3/3\pi^2$ where the Fermi wavevector $k_F =
    1/\delta x=1$.  The harmonic trapping potential along $z$ is then increased
    slowly while applying the quantum cooling algorithm described
    in~\cite{Bulgac:2013c} to cool the system to a state with two separated
    clouds.  These are the phase imprinted with $\delta\phi = \pi$ and the knife
    edge is removed, allowing the soliton to evolve as shown. Movies, including
    a case for a $48\times 48\times 128 $ lattice, may be found
    in~\cite{EPAPS:aps}.  This ring then oscillates along the axis of the trap.
    In the smaller simulation, the ring does not fully form, and it collapses in
    on itself, re-forming as a dark-soliton near the turning points.  This
    behavior mirrors that seen in \acs{BEC}~\cite{Shomroni:2009}, but is
    demonstrated here for the first time in a fermionic system.  This new domain
    wall exhibits the same initial instability, and a vortex ring of the
    opposite circulation and similar size forms and moves back along the trap in
    the opposite direction.  This oscillation is at the limit of the fermionic
    equivalent of the domain-wall branch of these types of
    excitations~\cite{Komineas:2007}. Note that~\cite{Komineas:2007} also
    discusses collisions of these excitations, which are elastic at low
    energies.  Reducing the width of the trap, one will continuously approach
    the quasi-\mysc{1D} situation of oscillating domain walls.  Note that the
    period $T \approx \sqrt{3}T_z$ in this case approximately agrees with other
    the quasi-\mysc{1D} simulations~\cite{Liao:2011, Scott:2011, Bulgac:2011c} }
\end{figure*}

This behaviour has been studied for bosons (see e.g.~\cite{Komineas:2007,
  Neely:2010, *Engels:2010}), where the transition from a domain wall to a
vortex ring appears to be continuous in harmonic tubes
(see~\cite{Komineas:2007}).  Our simulations shown in Fig.~\ref{fig:vortexring}
suggest that these results also apply qualitatively for fermions.  In
particular, in the smallest system (left panel of Fig.~\ref{fig:vortexring}),
the vortex ring configuration exists only away from the turning points.  It
collapses in on itself, re-forming as a domain wall near the turning points and
remerges as a vortex ring with an opposite circulation.  This behavior mirrors
that seen in \acs{BEC}~\cite{Shomroni:2009}, but is demonstrated here for the
first time in a fermionic system.  This new domain wall exhibits the same
initial instability, and a vortex ring of the opposite circulation and similar
size forms and moves back along the trap in the opposite direction.  This
oscillation is at the limit of the fermionic equivalent of the domain-wall
branch of these types of excitations~\cite{Komineas:2007}. Note
that~\cite{Komineas:2007} also discusses collisions of these excitations, which
are elastic at low energies.  Reducing the width of the trap, one will
continuously approach the quasi-\mysc{1D} situation of oscillating domain walls.
Note that the period $T \approx \sqrt{3}T_z$ in this case approximately agrees
with other the quasi-\mysc{1D} simulations~\cite{Liao:2011, Scott:2011,
  Bulgac:2011c}


The motion of a vortex ring in a trapped gas will be modified by the boundary:
the outward buoyant force of the trap, for example, will change the axial
velocity of the vortex ring according to the well established Magnus
relationship.  An oscillation can occur whereby a small vortex ring moves along
the axis of a trap, primarily according to~\eqref{eq:1} and the longitudinal
component of the buoyant force, then returns as a larger ring crawling along the
edge of the trap (see e.g.~\cite{Komineas:2007, Neely:2010, *Engels:2010}).
This picture follows from arguments similar to those used to
derive~\eqref{eq:vortex_ring}, but where the superflow outside the ring
no-longer extends to infinity.  As the boundary is reduced, the balance shifts
between the opposite flows inside and outside of the ring, slowing, then
ultimately reversing the velocity of the ring.  A quantitative analysis must
include effects such as the entrainment of the surrounding fluid, which can
change the effective mass, etc.; see~\cite{Magierski:2004a} for a few idealized
examples. Thus vortex rings can naturally oscillate with periods much larger
than $T_z$ in traps where the size of the transverse direction $R\gg 1/k_F$ is
large.

\paragraph{TDSLDA Model}
To demonstrate the generic generation of vortex rings, we simulate the
\ac{SLDA}~\eqref{eq:SLDA} on three \mysc{3D} lattices of size $24\times 24\times
96$, $32\times 32\times 128$, and $48\times 48\times128$ with unit lattice
spacing.  We adjust the particle number –– about $230$, $560$, and $1270$
particles for these lattices respectively –– so that the density in the core of
the initial cylindrical trap~\cite{Bulgac:2013c, EPAPS:aps} corresponds to $k_F
= 1$.  We evolve about $10^5$ wavefunctions in real time using a symplectic
split-operator integrator that respects time-reversal invariance using hundreds
of \acp{GPU} on the Titan supercomputer~\cite{titan}.  Preparing initial states
in \mysc{3D} has been a major challenge for superfluid \acs{DFT} like the
\ac{SLDA}, but the quantum-friction algorithm introduced in~\cite{Bulgac:2013c}
easily overcomes this challenge, and we quickly cool into the ground state of an
elongated harmonic trap. As in the experiment, we phase imprint a domain wall,
but to reduce phonon noise generated during the imprint, we include a repulsive
knife-edge potential.  Although the initial conditions have axial symmetry, the
simulations here are fully \mysc{3D} so as not to bias the results.

The simulations evolve a formally infinite system of coupled nonlinear
time-dependent \acs{PDE}s as described in detail elsewhere~\cite{Bulgac:2007a,
  *Bulgac:2011, *Bulgac:2013b, Bulgac:2011b}.  The \ac{TDSLDA} is based on the
simplest possible energy density functional that satisfies all expected
symmetries.  In addition to the number density $n = 2 \sum_{E_n < E_c}
\abs{v_n}^2$, the Pauli exclusion principle is ensured by including a kinetic
density $\tau_c = 2 \sum_{E_n < E_c} \abs{\vect{\nabla} v_n}^2$ in the spirit of
the original \ac{LDA} introduced by Kohn and Sham~\cite{Kohn:1965fk}, and
superfluidity is modelled by an anomalous density $\nu_c = \sum_{E_n < E_c}
v_n^{*} u_n$. Galilean covariance is restored by including the mass current
$\vect{j} = \tfrac{\hbar}{m} \sum_{E_n < E_c}2\Im v_n\vect{\nabla}v_n^*$ as
discussed in detail in~\cite{Engel:1975, Bulgac:2007a, *Bulgac:2011,
  *Bulgac:2013b}:%
\begin{multline}
  \mathcal{E} = \frac{\hbar^2}{m} \Biggl[
    \alpha \frac{\tau_c}{2} - (\alpha - 1)\frac{j^2}{2n}
    + \beta \frac{3 (3 \pi^2)^{2/3} n^{5/3}}{10} + \\
    + \frac{\abs{\nu_c}^2}{n^{1/3}/\gamma + \Lambda_c}
  \Biggr]
  + V_{\text{ext}}n, \label{eq:E_SLDA}
\end{multline}
(See Ref.~\cite{Bulgac:2007a, *Bulgac:2011, *Bulgac:2013b} for details on how to
express the regulator $\Lambda_c$ in terms of the energy cutoff $E_c$.)  

The correction to the energy density \eqref{eq:E_SLDA} for $\alpha \neq 1$ is
\begin{equation}
  (\alpha -1)\frac{\hbar^2}{2m} \left [ \tau_c -\frac{j^2}{n}\right ] = 
  (\alpha -1)\frac{\hbar^2}{2m} \left [ \tau_c -nv^2\right ],
\end{equation}
where $\vect{v}=\vect{j}/n$ is the local velocity.  The coefficient $(\alpha-1)
\approx 0.1$ in from of $\tau_c$ has the effect of slightly softening gradients;
the term proportional to $j^2$ is dominated by phase gradients and vanishes in
the ground state where there are no currents $\vect{j}=0$. Upon variation ,this
correction to the energy density leads to a change in the single-particle
Hamiltonian
\begin{multline}
  hv_k \rightarrow hv_k + (\alpha -1) \Biggl[
    -\frac{\hbar^2}{2m} \vect{\nabla}^2 v_k
    -2\I\hbar \vect{v}\cdot \vect{\nabla} v_k \\
  + \left(
      -\I\hbar\vect{\nabla} \cdot \vect{v} + \frac{m\vect{v}^2}{2}
    \right) v_k \Biggr].
\label{eq:Galilean}
\end{multline}
The gradient term proportional to local velocity $\vect{v}=\vect{j}/n $ acts
like a gauge potential, and will most affect mostly the flow, barely affecting
the density. The last term is a correction to the self-energy of the particle
and its magnitude, even close to a vortex core, is very small in the case of a
unitary Fermi gas, $\approx 0.1\times 0.2^2\varepsilon_F= 0.004 \varepsilon_F$,
since the maximum value of the velocity around a quantized vortex core is $v\leq
0.2v_F$~\cite{BY:2003}.

\begin{figure}[h]
  \includegraphics[width=\columnwidth]{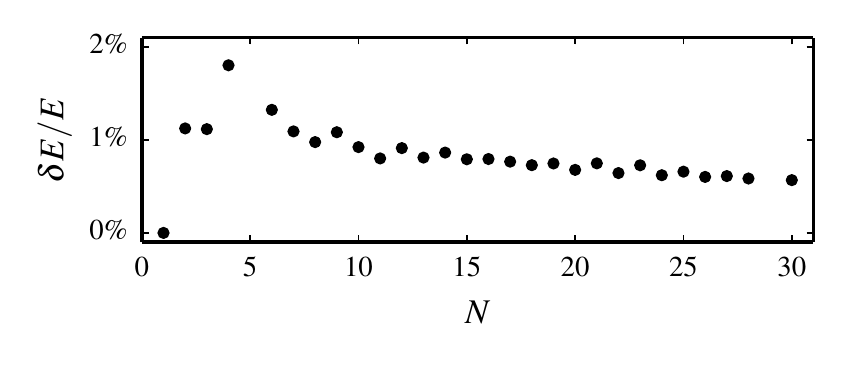}
  \caption{(color online) \label{fig:E_alpha}%
   Relative energy change (in \%) between simulations with $\alpha =
  1$ and $\alpha = 1.14$ that compute the energy of $N$ fermions (both even 
  and odd particle numbers) in the unitary Fermi
  gas trapped in an isotropic harmonic oscillator.  The parameters $\beta$ and
  $\gamma$ for the SLDA here have been adjusted to match the values of $\xi$ and
  $\eta$ used in Ref.~\cite{Bulgac:2007a} for fixed $\alpha = 1.14$ and
  $\alpha = 1.00.$  This demonstrates that a 14\% change in $\alpha$ results in less
  than 2\% change in energies, even in very small systems which have relatively
  large gradients. The use of the bare mass in \gls{DFT} calculations is in 
  the spirit of original Hohenberg and Kohn and Kohn and Sham formulation,
  which is widely used in electronic calculations in condensed matter physics 
  and chemistry.   }
\end{figure}

As with all \acp{DFT}, this can be applied to an arbitrary external one-body
potential $V_{\text{ext}}(\vect{x}, t)$.  In our first implementation of the
time-dependent GPU version of the code we use a split operator method which
requires less RAM memory.  For this method, gradient terms such as in
Eq.~\eqref{eq:Galilean} are extremely expensive to compute. In our simulations,
we use a simpler parameterization with $\alpha=1$ which avoids the need for the
current-dependent term on the second line of~\eqref{eq:Galilean}.  This affects
the quantitative accuracy of the theory only at the level of a few percent, see
also Fig.~\ref{fig:E_alpha} and the discussion below concerning gradient
corrections. The dynamical evolution is described by equations for the
quasiparticle wave functions $(u_k,v_k)$
\begin{equation}
  \I \hbar \pdiff{}{t}
  \begin{pmatrix}
     u_k \\ 
     v_k
   \end{pmatrix}
  =
  \begin{pmatrix}
    h  & \Delta \\
    \Delta^* & -h
  \end{pmatrix}
  \begin{pmatrix}
    u_k \\
    v_k
  \end{pmatrix},
  \label{eq:5}
\end{equation}
where the single-particle Hamiltonian $h$ and pairing potential $\Delta $ are
obtained by taking the appropriate functional derivatives of the energy density
$\mathcal{E}$.  The dimensionless constants $\alpha$, $\beta$, and $\gamma$ are
fixed by the energy per particle, pairing gap and quasiparticle spectrum
obtained from \ac{QMC} calculations of the homogeneous infinite system.  Though
Eq.~\eqref{eq:5} has a similar form as the mean-field \ac{BdG} equations, it
includes a self-energy contribution, which is dominant even at unitarity unlike
\ac{BdG}, and it includes all correlations at the same level of accuracy as the
\gls{QMC} results available so far.  

\begin{figure*}%
  \includegraphics[width=0.8\textwidth]{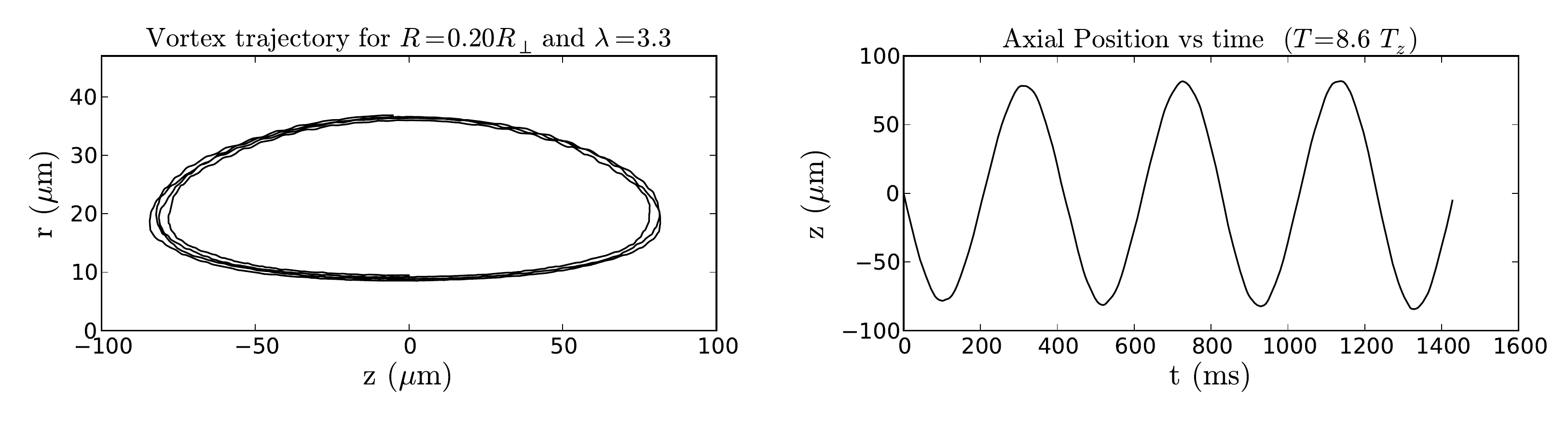}\\[-2em]
  \includegraphics[width=0.8\textwidth]{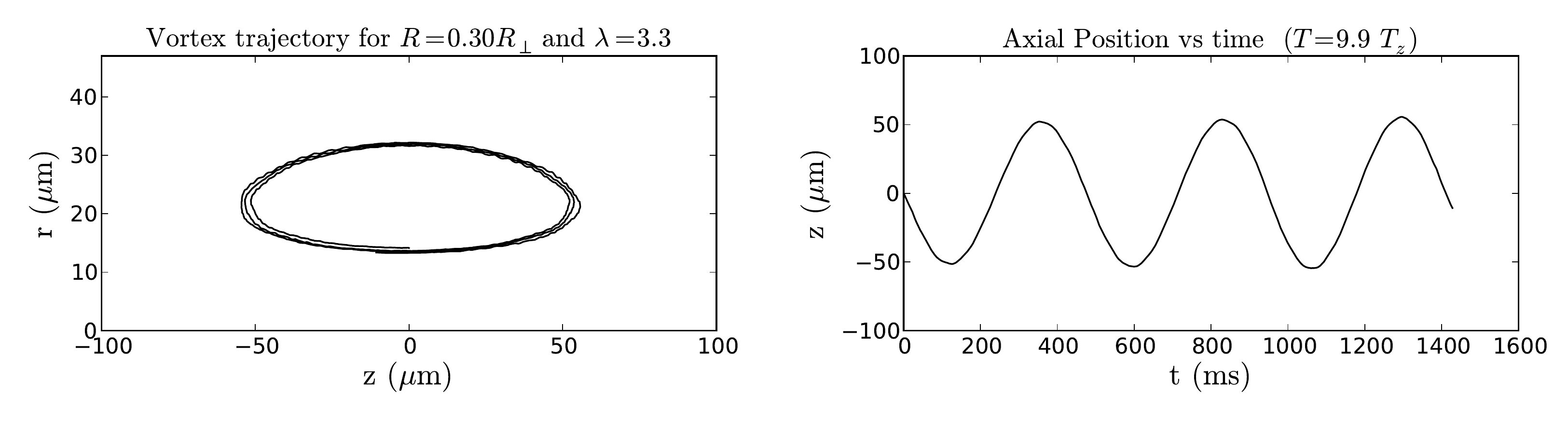}\\[-2em]
  \includegraphics[width=0.8\textwidth]{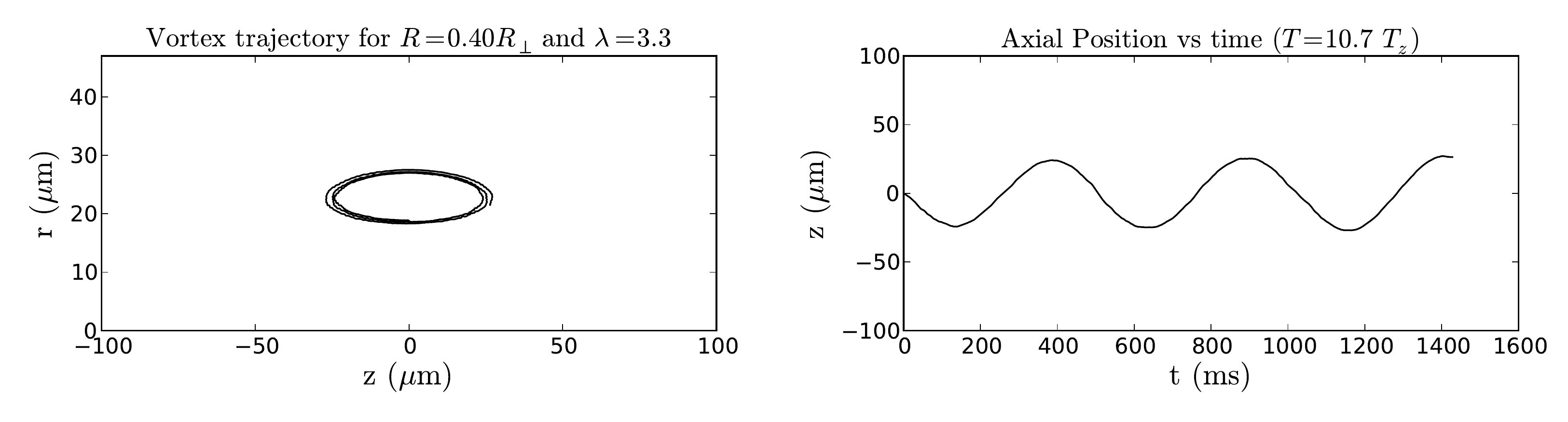}\\[-2em]
  \includegraphics[width=0.8\textwidth]{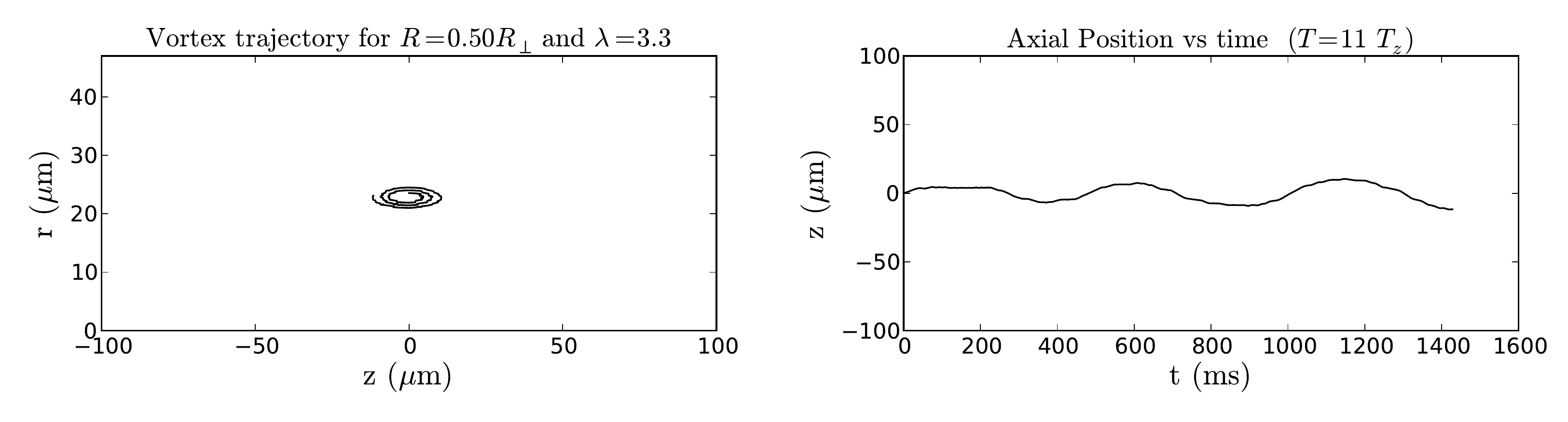}\\[-2em]
  \includegraphics[width=0.8\textwidth]{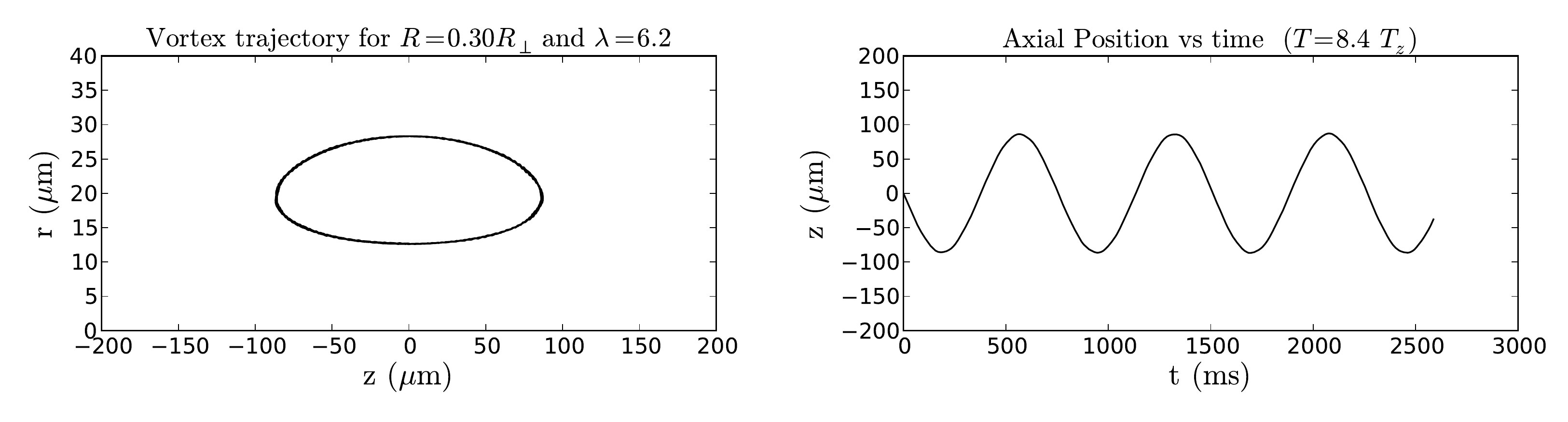}\\[-2em]
  \includegraphics[width=0.8\textwidth]{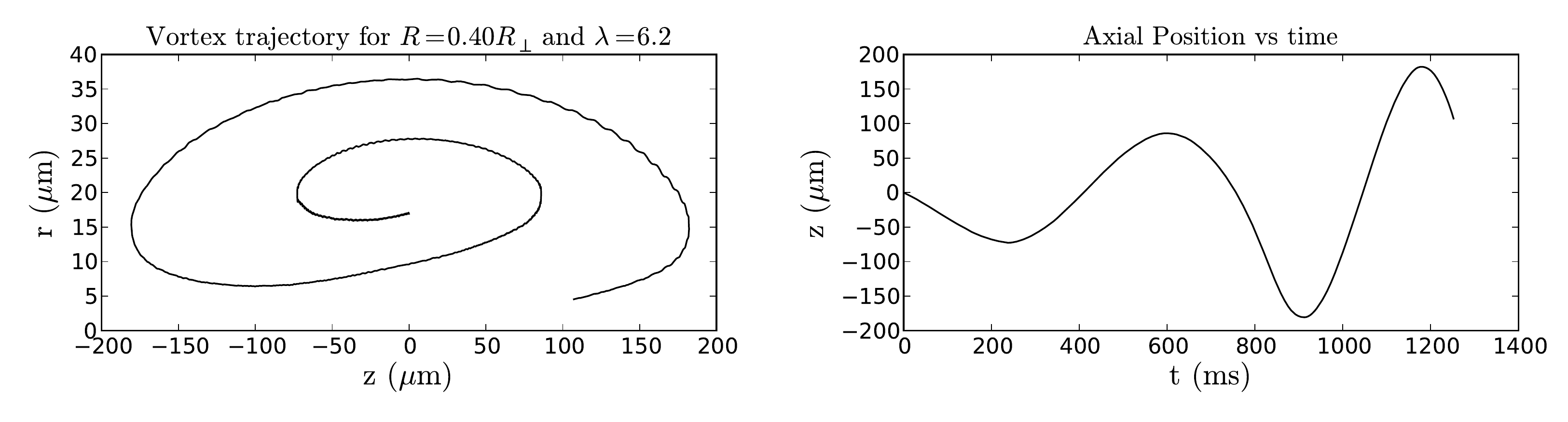}\\[-2em]
  \caption{\label{fig:ETF_summary}%
    Left column shows various trajectories of a vortex ring in the
    $R$-$z$--plane, while the right column shows the time dependence of
    corresponding $z$-coordinate of the vortex ring. The forth row show an
    example of an almost stationary vortex ring. The radius of a stationary
    vortex ring is $\approx 0.49R_\perp$, where $R_\perp$ is the \gls{TF} radius
    of the cloud. The last row shows an example of a vortex ring trajectory in
    the presence of a considerable number of phonons.}
\end{figure*}

\paragraph{ETF Model}
To demonstrate the scaling of oscillation periods with system size, aspect
ratio, etc.\@ we simulate the \gls{ETF} model~\eqref{eq:ETF} in a cylindrical
\gls{DVR} basis~\cite{LC:2002} with \num{2048} points along the $z$-axis and
\num{256} points in the radial direction, using trapping parameters as described
in~\cite{Yefsah:2013}.  We phase imprint a vortex ring with phase $\phi = \arg[z
+ \I(r - R)]$ ($r=\sqrt{x^2+y^2}$ and $R$ is the vortex ring radius) and
``cool'' with imaginary time evolution to generate vortex rings with various
amounts of background phonon excitations.  These are then evolved in real-time
using the split-operator integrator to determine the oscillation period, and to
perform the rapid-ramp/expansion imaging procedure.  Several sample results are
shown in Fig.~\ref{fig:ETF_summary}, and summarized in the following tables.

The \gls{ETF} follows from minimizing the energy-density
\begin{equation}
  \label{eq:E_ETF}
  \mathcal{E} = \frac{\abs{\vect{\nabla}\Psi}^2}{4m} +
  \mathcal{E}_{h}(n, a) + V_{\text{ext}}n,
\end{equation}
where $n = 2\abs{\Psi}^2$ is the total density and we parameterize the equation
of state of a dilute Fermi gas for positive scattering lengths with
\begin{equation}
  \mathcal{E}_h(n,a) = \frac{3}{5}\varepsilon_F n \xi
  \frac{\xi+x}{\xi+x(1+\zeta)+3.0\pi \xi x^2} 
  -\frac{\hbar^2}{2ma^2}n
\end{equation}
which depends on the magnetic field $B$ through the two-body scattering length
$a$ as described in~\cite{Bartenstein:2005uq, Zurn:2013} through the
dimensionless interaction parameter $x=1/k_Fa$.  Dimensions are set in terms of
the parameters of the free Fermi gas $n = k_F^3/3\pi^2$, and
$\varepsilon_F=\hbar^2k_F^2/2m$. This reproduces the unitary equation of state
with $\xi=0.370$ and $\zeta=0.901$ (the contact), and the factor $3.0 =
9a/5a_{DD}$ reproduces the dimer-dimer scattering length $a_{DD} \approx 0.6
a$. The \gls{ETF} approach that has been used to analyze the
expansion~\cite{Kim:2004a} and~\cite{Kim:2004a, *Kim:2004, *Kim:2005,
  *Kim:2005a} breathing mode frequencies of cold atomic gases in a trap, their
surface oscillations~\cite{Salasnich:2010, *Salasnich:2012}, collisions of
clouds of fermions~\cite{Ancilotto:2012, *Ancilotto:2012a}, vortex
generation~\cite{Ancilotto:2013}, vortex pinning~\cite{Bulgac:2013a}, and
soliton dynamics~\cite{Khan:2013}.  The \gls{ETF} is equivalent to the quantum
hydrodynamics approach at zero temperature, which has been used extensively by
many authors for modelling the unitary Fermi gas during the last decade, but
\gls{ETF} also includes the quantum pressure term typically neglected in a
hydrodynamic approach.  We point out that the \gls{ETF} in Eq.~\eqref{eq:ETF} is
manifestly covariant under Galilean transformations whereby $\Psi(\vect{x}, t)
\rightarrow e^{-2\I\phi}\Psi(\vect{x} + \vect{v}t, t)$ where $\phi =
m\vect{v}\cdot\vect{x} + \tfrac{1}{2}mv^2t$.  The factor $2$ in the exponent
here corresponds with the identification of $\Psi = \braket{\psi_a\psi_b}$ as
the dimer or di-fermion condensate and could be absorbed into the definition of
the dimer mass $m_B = 2m$.

\paragraph{Gradient corrections}
The \gls{SLDA}~\eqref{eq:E_SLDA} and \gls{ETF}~\eqref{eq:E_ETF} functionals
naturally describe gradients through their kinetic terms, but since they have
been fit to properties of homogeneous system, one might reasonably wonder if any
significant gradient corrections have been overlooked.  This question was
addressed in~\cite{Forbes:2012a} by confronting high-precision \gls{QMC}
calculations of harmonically trapped systems which found that, while small
gradient corrections of Weizsäcker-type $\propto -\abs{\vect{\nabla} n}^2/n$
might improve the ability of the functionals to fit traps with few particles,
they can be treated as a perturbative correction to the underlying \gls{SLDA}
\eqref{eq:E_SLDA}.  Indeed, the \gls{SLDA} performs extremely well, even when
applied to inhomogeneous trapped systems, where it explains virtually all
available \gls{QMC} results for systems with up to 120 fermions, both in
polarized and unpolarized systems, in superfluid or normal phases, in harmonic
traps, periodic boxes, and infinite matter~\cite{Bulgac:2007a, *Bulgac:2011,
  *Bulgac:2013b, FGG:2010, *Forbes:2012}.  This indicates that the kinetic
energy density $\tau_c = 2\sum_{E_n < E_c} \abs{\vect{\nabla} v_n}^2$ properly
describes most of the gradient effects.  Since the pairing is strong in the
unitary Fermi gas, this kinetic energy density includes gradients up to high
momenta $\gg k_F$, much larger than the scales $\sim k_F$ describing solitons
like domain-walls, vortices, and vortex rings.  Thus, though additional gradient
corrections may affect the structure of solitons, these effects will be subtle,
and will require high-precision \gls{QMC} calculations and measurements to
validate.  The omission of these higher-order corrections in our simulations are
unlikely to affect our results more than the quoted experimental accuracy, as
also illustrated in Fig.\ref{fig:E_alpha}, for a case where density gradients
are significant.

The available \gls{QMC} calculations are not accurate enough to provide a more
exact value of the effective mass of the quasi-particles.  The \gls{QMC} results
of~\cite{Carlson:2005kg} suggest a value of $\alpha \approx
1.14$~\cite{Bulgac:2007a} obtained by calculating the ground state energy
difference $\abs{E(N\pm 1,k) - E(N,0)}$, where $N$ is even and $\hbar k$ is the
momentum of the system with $N\pm 1$ particles.  In that calculation the Bertsch
parameter was determined to be $\xi=0.42(2)$, thus with an error of about 10\%
when compared to more recent \gls{QMC} results~\cite{Carlson:2011} and
experimental measurements $\xi=0.372$~\cite{Ku:2011, Zurn:2013}.  One can
therefore infer that the error in the quasi-particle dispersion is much larger
as it is a difference in energies.  The \gls{QMC}
calculation~\cite{Carlson:2005kg} for the pairing gap quotes an error of about
5\% ($\Delta =0.504(24)\varepsilon_F$).  This agrees within the error bars with
an independent analysis of experimental data of polarized
system~\cite{Carlson;Reddy:2008-04}, which claims that $\Delta =
0.45(5)\varepsilon_F$.  In an independent \gls{QMC} study at finite temperatures
~\cite{Magierski:2009}, the effective mass was found to be consistent with the
bare mass to within 10\% for a large range of temperatures. In summary, all
direct information from \gls{QMC} calculations and experiments contain errors at
the level of about 10\%, statistical as well systematic, consistent with
$\alpha-1\leq 0.1$.

\begin{figure}[htb]
  \includegraphics[width=\columnwidth]{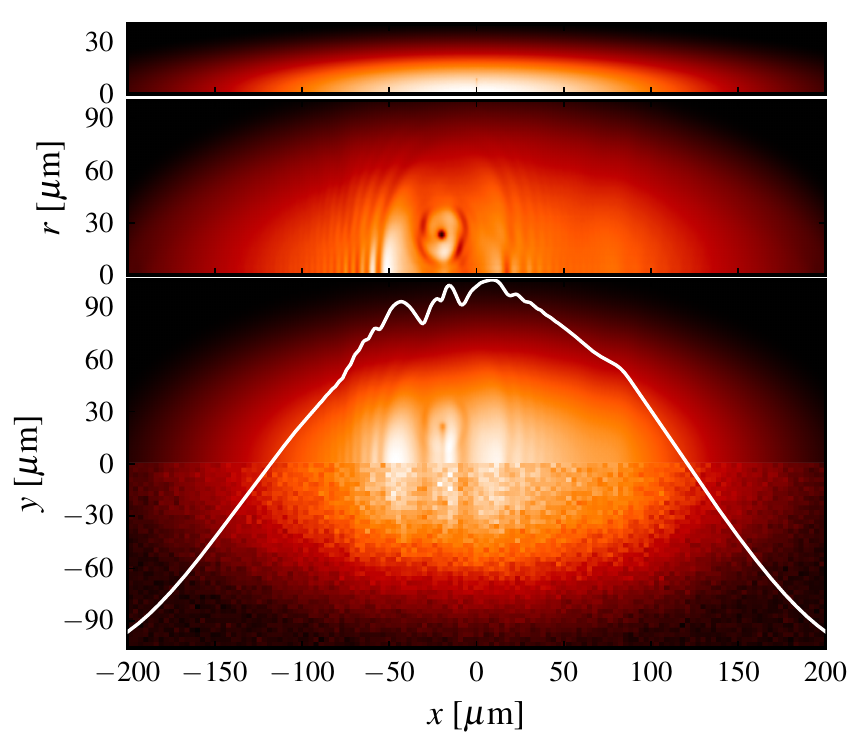}
  \includegraphics[width=\columnwidth]{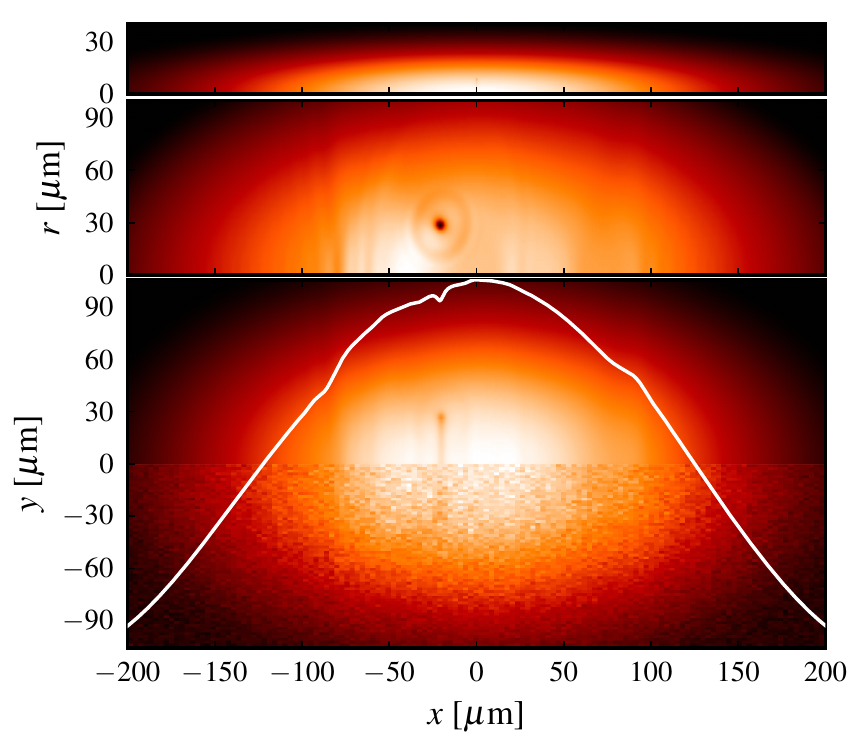}
  \caption{(color online) \label{fig:imag_2}%
    Additional expansion images with same interpretation as in
    Fig.~\ref{fig:imag}. Top: image of a small vortex ring expanded at
    $B_{\text{min}}=580$~G. Bottom: image of same small vortex ring expanded at
    $B_{\text{min}}=702$~G demonstrating that $B_{\text{min}}<700$~G is required
    to achieve a resolvable image, thereby explaining the need for the subtle
    time-of-flight expansion and imaging procedure discussed in the
    experiment~\cite{Yefsah:2013}. See~\cite{EPAPS:aps} for movies.}
\end{figure}

\paragraph{Results}
The results of the expansion and imaging process are shown in
Fig.~\ref{fig:imag} and Fig.~\ref{fig:imag_2} (see~\cite{EPAPS:aps} for movies).
Here it becomes clear why the experiment~\cite{Yefsah:2013} needed to implement
an involved ramping/imaging procedure to image the defects.  The rapid ramp into
the \gls{BEC} regime at low magnetic fields $B_{\text{min}} < 700$~G causes a
rapid change in the coherence length that produces a sort of shock-wave during
the expansion.  The geometry to the expansion results in an asymmetric density
depletion that resolves into a planar looking object when imaged with the
coarse-grain resolution of the imaging system.  If the field $B_{\text{min}}$
not sufficiently small, then the shock-wave is mild, and the resulting cloud
does not have enough contrast to register a signal.  That the \gls{ETF}
quantitatively reproduces the required minimum field $B_{\text{min}}$ is further
validation of the \gls{DFT} and somewhat expected since the \gls{ETF}
approximation should become more accurate in the \gls{BEC} regime.

The \gls{ETF} certainly cannot reproduce all details of the fermionic dynamics
-- in particular, one expects poor behavior when excitations approach the
pair-breaking threshold set by the gap $\hbar\omega > 2\Delta \approx E_F$.  The
theory, however, has the same symmetries, and is tuned to have the same equation
of state as the full theory. The advantage of this approach over traditional
fermionic \glspl{TDDFT} is its computational simplicity: the bosonic approach
needs only to evolve a single wavefunction.  A detailed comparison of the
\gls{ETF} and \gls{SLDA} is performed in~\cite{Forbes:2012b}.

In Table~\ref{tab:ETF_aspect} we compared the oscillation periods predicted by
the \gls{ETF} with the observed period from~\cite{Yefsah:2013} on resonance for
the three aspect ratios studied in the experiment.  The observations are
consistently larger than the \gls{ETF} predictions by a factor of about $1.8$:
this might be due to the lack of a normal component filling the core of the
vortices in the \gls{ETF} and is reminiscent of the factor $\sqrt{3/2}$
difference in the calculated period of \mysc{1D} domain walls. This is
consistent with the heuristic estimate~\eqref{eq:ring_period} whereby the mass
depletion $M_{\textit{VR}}$ would be suppressed for fermions by the presence of
the normal state.

To test the consistency of this suppression, we use the \gls{ETF} to model the
\gls{TDSLDA} simulations shown in Fig.~\ref{fig:vortexring}.  The comparison is
shown in Table~\ref{tab:ETF_SLDA} where it is seen that the \gls{TDSLDA} periods
are larger than the \gls{ETF} by a factor consistent with $\sqrt{3/2}$ for small
systems.  This is expected since these simulations are in small traps and are
very close to the limit where domain walls remain stable. The lattice simulation
$48\times48\times128$, which involved \num{259762} complex time-dependent
\mysc{3D} nonlinear coupled partial differential equations, performed on
Titan~\cite{titan} on \num{2048} GPUs, is one of the largest Direct Numerical
Simulations (DNC) performed so far.

\begin{table}[H]
  \caption{\label{tab:ETF_SLDA}%
    Benchmark of the \gls*{ETF} periods to the \gls*{SLDA} periods for sizes
    $24\times24\times 96$, $32\times32\times128$, and $48\times48\times128$.}
  \begin{ruledtabular}
    \begin{tabular}{cccc}
      Size & $T_\text{ETF}$ & $T_{\text{SLDA}}$
      & $T_{\text{SLDA}}/T_{\text{ETF}}$\\ 
      \hline         
      $24\times24\times96$ & $1.4 T_z$ & $1.7 T_z$ & 1.2\\
      $32\times32\times128$ & $1.6 T_z$ & $1.9 T_z$ & 1.2\\
      $48\times48\times128$ & $1.9 T_z$ & $2.6 T_z$ & 1.4\\
    \end{tabular}
  \end{ruledtabular}
\end{table}

In Table~\ref{tab:ETF_radius} we demonstrate how the period depends on the
initial radius of the imprinted vortex ring $R_0$.  This parameter is not
directly measured or controlled in the experiment, so we must estimate the value
$R_0 \approx 0.2R_\perp$ by the resulting amplitude of oscillation $\sim 0.5R_z$
shown in the figures of~\cite{Yefsah:2013}.
\begin{table}[H]
  \caption{\label{tab:ETF_radius}%
    Imprinting the vortex with different radii, all on resonance with
    $1/\lambda=3.3$. In the tables below we show how oscillation period changes
    with aspect ratio for a vortex ring imprinted at $R_0=0.30~ R_\perp$ on
    resosnace $k_Fa = \infty$ in each scenario.}
  \begin{ruledtabular}
    \begin{tabular}{lll}
      Imprint radius & Period & Amplitude\\               
      \hline
      $R_0=0.20~R_\perp$ & $T=\hphantom{1}8.6~T_z$ & $\sim0.45~R_z$\\
      $R_0=0.30~R_\perp$ & $T=\hphantom{1}9.9~T_z$ & $\sim0.35~R_z$\\
      $R_0=0.40~R_\perp$ & $T=10.7~T_z$ & $\sim0.15~R_z$\\
      $R_0=0.50~R_\perp$ & $T=11.0~T_z$ & $\sim0.05~R_z$
    \end{tabular}
  \end{ruledtabular}
\end{table}

Finally, we comment on the observed ``snake'' instability discussed in the
supplementary information of~\cite{Yefsah:2013}.  Although significantly more
stable than domain walls, large vortex rings can also bend and decay through the
Crow instability~\cite{Crow:1970, Gautam:2012} and the \gls{MIT} experiment is
poised right on the edge of the regime where one can start to explore the
quantum turbulence cascade.
   

%

\end{document}